\documentclass[reqno]{amsart}
\newtheorem{thm}{Theorem}[section] 
\newtheorem{cor}[thm]{Corollary} \newtheorem{lem}[thm]{Lemma} 
 \theoremstyle{definition} 
\newtheorem{defn}[thm]{Definition} \theoremstyle{remark} 
\newtheorem{rem}[thm]{Remark} 
 
\numberwithin{equation}{section} 
\usepackage{amsfonts}
 \usepackage{amsmath}
 \usepackage{amssymb}
\usepackage{graphicx}
\usepackage{epsfig}
\usepackage{color}
\usepackage{amsmath,amscd,amssymb,amsfonts}
 
\begin{document}
\title[ Scattering problems for symmetric systems with dissipative boundary conditions] {Scattering problems for symmetric systems with dissipative boundary conditions }
\author{Vesselin Petkov}\address{Institut de Math\'ematiques de Bordeaux, 351 Cours de la Lib\'eration, 33405 Talence, France}\email{petkov@math.u-bordeaux1.fr}
\date{}
\newcommand{\h}{{\mathcal H}}
\newcommand{\R}{{\mathbb R}}
\newcommand{\N}{{\mathbb N}}
\newcommand{\C}{{\mathbb C}}
\newcommand{\F}{{\mathcal F}}
\newcommand{\Oo}{{\mathcal O}}
\newcommand{\K}{{\mathcal K}}
\newcommand{\D}{{\mathcal D}}
\newcommand{\G}{{\mathcal G}}
\newcommand{\Hh}{{\mathcal H}}
\newcommand{\Z}{{\mathbb Z}}
\newcommand{\Q}{{\mathbb Q}}
\newcommand{\U}{{\mathcal U}}
\newcommand{\A}{{\mathbb A}}
\newcommand{\Ss}{{\mathbb S}}
\renewcommand{\Re}{\mathop{\rm Re}\nolimits}
\renewcommand{\Im}{\mathop{\rm Im}\nolimits}

\maketitle

\numberwithin{equation}{section}

\def\CC{{\mathcal C}}
\def\lap{\bigtriangleup}
\def\ra{\rangle}
\def\la{\langle}
\def\TO{\dot{T}^*(\Omega)}
\def\ff{{\mathcal F}}
\def\rr{{\mathcal R}}
\def\ot{(\omega, \theta)}
\def\e{\varepsilon}
\def\phi {\varphi}
\def \la {{\lambda}}
\def \a {{\alpha}}
\def\sn{{\mathbb S}^{n-1}}
\def\ssn{{\sn}\times {\sn}}
\def\pp{P_{+}}
\def\ppm{P_{-}}
\def\ppr{P_{+}^{\rho}}
\def\ppmr{P_{-}^{\rho}}
\def\hl{{\mathcal H}_{{\rm loc}}}
\def\rp{{\rm Res}\:P(t)}
\def\ggi{{\mathcal G}^{(i)}}
\def\td{\tilde{d}}
\def\cS{\check{S}}
\def\ii{{\mathcal I}}
\def\mm{{\mathcal M}}
\def\cp{\check{p}}
\def\pr{{\rm pr}}
\def\uu{{\mathcal U}}
\def\qq{{\mathcal Q}}
\def\ovo{\overset{\circ}\Omega}
\def\ii{{\bf i}}
\def\la{\langle}
\def\ra {\rangle}
\def\pt{P_{\theta}}
\def\Rc{R_{\chi}(\lambda)}
\def\Ha{H_0^{ac}}
\def\Hi{{\mathcal H}_{\infty}}
\def\Rc{{\mathcal R}}
\def\curl{{\rm curl}\,}
\def\dive{{\rm div}\,}
\def\grad{{\rm grad}\,}
\def\dk{\partial K}
\def\cn{{\mathcal N}}
\def\d{{\partial}}
\def\mc{{\mathcal H}}
\def\ssn{\mathbb S^{n-1} \times \mathbb S^{n-1}}
\def\ii{{\bf i}}
\def\la{\langle}
\def\ra{\rangle}
\def\sn{{\mathbb  S}^{n-1}}
\def\Dp{D_{+}^{\rho}}
\def\Dm{D_{-}^{\rho}}
\def\Dpm{D_{\pm}^{\rho}}
\def\Ker{{\rm Ker}\:}
\def\Im{{\rm Im}\:}

\begin{abstract} We study symmetric systems with dissipative boundary conditions. The solutions of the mixed problems for such systems are given by a contraction semigroup $V(t)f = e^{tG_b}f,\: t \geq 0$ and the solutions $u = e^{tG_b}f$ with eigenfunctions $f$ of the generator $G_b$ with eigenvalues $\lambda,\: \Re \lambda < 0,$ are called asymptotically disappearing (ADS). We prove that the wave operators are not complete if there exist (ADS). This is the case for Maxwell system with special boundary conditions in the exterior of the sphere. We obtain a representation of the scattering kernel and we examine the inverse back-scattering problem related to the leading term of the scattering kernel.
\end{abstract}
\vspace{0.5cm}



\section{Introduction}
\renewcommand{\theequation}{\arabic{section}.\arabic{equation}}
\setcounter{equation}{0}

 Let $K \subset \{ x \in \R^n,\: |x| \leq \rho\},\: n \geq 3$, $n$ odd,  be an open bounded domain with $C^{\infty}$ boundary $\partial K.$ Consider in $\Omega = \R^n \setminus \bar{K}$ the operator $G = \sum_{j=1}^n A_j(x) \partial_{x_j} + B(x),$  where $A_j(x),\: j =1,...,n,$ are smooth symmetric $(r \times r)$ matrices and $B(x)$ is a smooth $(r \times r)$ matrix. For simplicity in this paper we will assume that $A_j$ are constant matrices and $B = 0$ but our results remain true for operators with variable coefficients under some conditions on the decay of $A_j(x)$ and $B(x)$ as $|x| \to \infty.$

We assume that the eigenvalues of $A(\xi) = \sum_{j = 1}^n A_j \xi_j$  for $\xi = (\xi_1,...,\xi_n) \in \R^n \setminus \{0\}$ have constant multiplicity independent of $\xi.$ Denote by $\nu(x) = (\nu_1(x),...,\nu_n(x))$ the unit normal at $x \in \partial \Omega$ pointing into
$K$ and set $A(\nu(x)) = \sum_{j=1}^n A_j \nu_j(x)$. Let $\cn(x) \subset \C^r$ be a linear space depending smoothly on $x \in \partial \Omega$ such that

$(i)\:\:\:   \langle A(\nu(x)) u(x), u(x) \rangle \leq 0$ for all $u(x) \in \cn(x),$

$(ii) \:\:\:  \cn(x)$ is maximal with respect to (i).

Consider the boundary problem
\begin{equation} \label{eq:1.1}
\begin{cases} (\partial_t - G) u = 0 \:{\rm in}\: \R^+ \times \Omega,\\
u(t,x) \in \cn(x) \: {\rm for}\: t\geq 0,\:x \in \partial \Omega,\\
u(0, x) = f(x) \: {\rm in}\: \Omega. \end{cases}
\end{equation}

The conditions (i), (ii) make it possible to introduce a contraction semigroup $V(t) = e^{tG_b},\: t \geq 0$ in $H = L^2(\Omega: \C^r)$ related to the problem (\ref{eq:1.1}) with generator $G_b$. The domain $D(G_b)$ of $G_b$ is the  closure with respect to the graph-norm $(\|g\|^2 + \|Gg\|^2)^{1/2} $ of functions $g(x) \in C_{(0)}^1(\bar{\Omega}: \C^r)$ satisfying the boundary condition $g(x)\vert_{\partial \Omega} \in \cn(x)$.\\
 
 Next consider the unitary group $U_0(t) = e^{t G_0}$ in $H_0 = L^2(\R^n: \C^r)$ related to the Cauchy problem
\begin{equation} \label{eq:1.2}
\begin{cases} (\partial_t - G) u = 0 \:{\rm in}\: \R \times \R^n,\\
u(0, x) = f(x) \: {\rm in}\: \R^n, \end{cases}
\end{equation}
where $G_0$ with domain $D(G_0) = \{f \in H_0:\: Gf \in H_0\}$ is the generator of $U_0(t).$
Let $H_b \subset H$ be the space generated by the eigenvectors of $G_b$ with eigenvalues $\mu \in \ii\R$ and let $H_b^{\perp}$ be the orthogonal complement of $H_b$ in $H.$ The generator $G_0 =  \sum_{j=1}^n A_j \partial_{x_j}$ is skew self-adjoint in $H_0$ and the spectrum of  $G_0$ on the space $\Ha = ({\Ker}\: G_0)^{\perp} \subset H_0$ is absolutely continuous (see Chapter IV in \cite{P}).\\

For dissipative symmetric systems some solutions can have  global energy decreasing exponentially as $t \to \infty$ and it is possible also to have disappearing solutions. The precise definitions are given below.

\begin{defn} 
We say that $u = V(t)f$ is a disappearing solution (DS), if there exists $T > 0$ such that $V(t)f = 0$ for $t \geq T.$
\end{defn} 

\begin{defn} 
We say that $u = V(t)f$ is asymptotically disappearing solution (ADS), if there exists $\lambda \in \C$ with $\Re \lambda < 0$ and $f \not= 0$ such that $V(t)f = e^{\lambda t}f$.
\end{defn}
Notice that if $V(t)f = e^{\lambda t}f,$ then $f \in D(G_b)$ and $G_b f= \lambda f.$ The existence of disappearing solutions perturb strongly the inverse back-scattering problem since the leading term of the back-scattering matrix vanishes for all directions (see Section 5). This phenomenon is well known for the wave equations with dissipative boundary conditions \cite{Ma4}, \cite{GA}, \cite{P}. For symmetric systems with dissipative boundary conditions the situation is much more complicated. It seems rather difficult to construct disappearing solutions and even for Maxwell system the problem of the existence of disappearing solutions remains open (see \cite{CPR}, \cite{P} and the references given there). In this paper  we present a survey of some results related to the existence of (ADS). First in Section 2 we show that the completeness of the wave operators $W_{\pm}$ related to $V(t)$ and $U_0(t)$, fails if (ADS) exist. Therefore we must define the scattering operator by using another operator $W.$  Secondly, we describe in Section 3 some recent results obtained in \cite{CPR}, where (ADS) for Maxwell system have been constructed. We study maximally dissipative boundary conditions for which there are no disappearing solutions but (ADS) exist. This shows the importance of (ADS) which are stable under perturbations. Next in Section 4 we establish a representation of the scattering kernel of the scattering operator in the case of characteristics of constant multiplicity following the arguments in \cite{MT}, \cite{P2} for strictly hyperbolic systems. Finally, in Section 5 we study the inverse back-scattering problem connected with the leading singularity of the scattering matrix $\Bigl(S^{j k}(s, -\omega, \omega)\Bigr)_{j,k=1}^d.$ Here the boundary condition plays a crucial role and we investigate the problem assuming that 
$$ \cn(x) \ominus \Ker (A(\nu(x)))\not= \Sigma_{-}(\nu(x)),$$
$\Sigma_{-}(\nu(x))$ being the space spanned by the eigenvectors of $A(\nu(x))$ with negative eigenvalues. This condition guarantees that at least one element of the scattering matrix $S^{j k}(s, -\omega, \omega)_{j,k=1}^d$ has a non vanishing leading singularity related to the support function $\rho(\omega) = \min_{x \in \partial \Omega} \la x, \omega \ra$ in direction $\omega \in \sn.$

\section{Wave operators}
\renewcommand{\theequation}{\arabic{section}.\arabic{equation}}
\setcounter{equation}{0}

To introduce the wave operators, consider the operator $J:\: H \longrightarrow H_0$ extending $f \in H$ as 0 for $x \in K$ and let $J^*:\: H_0 \longrightarrow H$ be the adjoint of $J.$ Let $P_{ac}$ be the orthogonal projection on the space $\Ha.$ The wave operators related to perturbed and non-perturbed problems have the form
$$W_{-}f = \lim_{t \to +\infty}  V(t) J^* U_0(-t) P_{ac} f,\: f \in H_0,$$
$$W_{+} f = \lim_{t \to +\infty} V^*(t) J^* U_0(t) P_{ac}f,\: f \in H_0.$$

Under the above hypothesis it is not difficult to prove the existence of $W_{\pm}$ and to show that (see for instance, \cite{GS} and Chapter III in \cite{P})
$${\rm Ran}\: W_{\pm} \subset H_b^{\perp}.$$

To obtain more precise results for Ran $W_{\pm}$, we need to impose the following {\it coercive conditions}.

$(H):\:$ For each $f \in D(G_b) \cap (\Ker G_b)^{\perp}$ we have 
$$\sum_{j=1}^n \|\partial_{x_j} f\| \leq C (\|f\| + \|G_bf\|)$$
with a constant $C > 0$ independent of $f.$ \\
 
$(H^*):\:$ For each $f \in D(G_b^*) \cap (\Ker G_b^*)^{\perp}$ we have
$$\sum_{j=1}^n \|\partial_{x_j} f\| \leq C (\|f\| + \|G_b^*f\|)$$
with a constant $C > 0$ independent of $f.$

\begin{rem} The conditions $(H)$ and $(H^*)$ are satisfied for a large class of non elliptic symmetric systems (see \cite{Ma2}) for which it is possible to construct a first order $(l \times r)$ matrix operator $Q = \sum_{j = 1}^n Q_j \partial_{x_j}$ so that 
$$Q(\xi) A(\xi) = 0,\:\: \Ker Q(\xi) = \Im A(\xi),$$
where $Q(\xi) = \sum_{j= 1}^n Q_j \xi.$
In our case we need coercive estimates for $f \in D(G_b) \cap H_b^{\perp}.$ On the other hand, the space $\Ker G_b$ is infinite dimensional if $Q$ with the properties above exist. Notice also that for the generator $G_b$ we have $\Ker G_g = \Ker G_b^*.$

\end{rem}

Introduce the spaces
$${\mathcal H}_{\infty}^+ = \{f\in H:\: \lim_{t \to + \infty} V(t)f = 0\}, \:{\mathcal H}_{\infty}^- = \{f\in H:\: \lim_{t \to + \infty} V^*(t)f = 0\}.$$

We have the following
\begin{thm} [\cite{G2}] Assume the conditions $(H)$ and $(H^*)$ fulfilled. Then
$$\overline{{\rm Ran}\: W_{\pm}} = H_b^{\perp} \ominus {\mathcal H}_{\infty}^{\pm}.$$ 
\end{thm}
To obtain a completeness of the wave operators we must have ${\mathcal H}_{\infty}^{+} = {\mathcal H}_{\infty}^{-}.$
On the other hand, in general, it is not clear if these subspaces are not empty.\\

For our analysis we use the translation representation ${\mathcal R}_n: \Ha \longrightarrow (L^2(\R \times \sn))^d$ of $U_0(t),$ where Rank $A(\xi) = r - d_0 = 2d > 0$ for $\xi \not= 0.$ Let $\tau_j(\xi),\: j = 1,...,2d,$ be the non-vanishing eigenvalues of $A(-\xi)$ ordered as follows
$$\tau_1(\xi) > ...> \tau_d(\xi) > 0 > \tau_{d+1}(\xi) > ...> \tau_{2d}(\xi),\: \xi \not= 0.$$
Denote by $r_j(\xi), \: j = 1,...,2d,$ the normalized eigenvectors of $A(-\xi)$ related to $\tau_j(\xi)$. Then $\Rc_n$ has the form (see Chapter IV in \cite{P})
$$(\Rc_n f)(s, \omega) = \sum_{j=1}^d \tilde{k}_j(s, \omega) r_j(\omega ),$$
where 
$$\tilde{k}_j(s, \omega) = \tau_j(\omega)^{1/2} k_{j}(s \tau_j(\omega), \omega),\: j = 1,...,d,$$
and $k_j(s, \omega) = 2^{-(n-1)/2}D_s^{(n-1)/2} \langle (Rf)(s, \omega), r_j(\omega) \rangle$, $(Rf)(s, \omega)$ being the Radon transform of $f(x).$ $\Rc_n$ map $\Ha$ isometrically into $(L^2(\R \times \sn))^d$ and $\Rc_n U_0(t) = T_t \Rc_n,\: \forall t \in \R$, where $T_t g = g(s- t, \omega).$ Let 
$0 < v_0 =  \min_{\omega \in \sn} \tau_d(\omega).$ To introduce the Lax-Phillips spaces (see Chapter VI, \cite{LP}) , we need the following
\begin{defn} We say that $f \in D_{\pm}$ if 
$$U_0(t)f = 0 \: {\rm for}\: |x|< \pm v_0 t,\: \:\pm t > 0.$$
\end{defn}
We have $f \in D_{\pm}$ if and only if $\Rc_n f(s, \omega) = 0$ for $\mp s > 0.$
Set $D_{\pm}^{b} = U_0(\pm b/v_0) D_{\pm},\: b > 0.$ For $t \geq 0$ it is easy to prove (see Lemma 4.1.5 in \cite{P}) the following equalities
$$U_0(-t) V(t) f = V^*(t)U_0(t)f = f,\: f \in \Dp,$$
$$U_0(t)V^*(t)f = V(t)U_0(-t)f = f,\: f \in \Dm.$$

The next result is similar to that in \cite{G3} established for strictly hyperbolic systems.

\begin{thm} If $f \in D(G_b^j) \cap {\mathcal H}_{\infty}^{+} \cap (D_{-}^{\rho})^{\perp}, \forall j \in \N, f = G_b f_0, f_0 \in H_{b}^{\perp}$, then we have $(V(t)f)(t, x) = 0$ for $|x| > 2 \rho.$
\end{thm}
This yields the following 
\begin{cor}  Assume that there exists an asymptotically disappearing solution $u(t,x) = e^{\lambda t}f(x)$ such that $f(x) \not= 0$ does not have a compact support. Then ${\mathcal H}_{\infty}^{+} \not= {\mathcal H}_{\infty}^{-}$ and the wave operators are not complete.
\end{cor}
{\it Proof.} If $G_b f = \lambda f,\: \Re \lambda < 0$, we have $ f \in D(G_b^j) \cap {\mathcal H}_{\infty}^{+},\: \forall j \in \N$. Assuming ${\mathcal H}_{\infty}^{-} = {\mathcal H}_{\infty}^{+}$, for $g \in \Dm$ we get
$$(f, g) = (f, V(t) U_0(-t) g) = (V^*(t)f, U_0(-t)g) \longrightarrow_{t \to \infty} 0.$$
Thus $f \in (\Dm)^{\perp}$ and we can apply Theorem 2.4.\\

{\it Proof of Theorem $2.4.$} Given $g \in {\mathcal H}_{\infty}^{+} \cap (\Dm)^{\perp}$ and $f \in \Dp$, for every fixed $t \geq 0$ we get
$$(V(t)g, f) = (V(t)g, V^*(s)U_0(s)f) = (V(t + s)g, U_0(s)f) \longrightarrow 0 \: {\rm as}\: s \to +\infty.$$
Thus, $V(t)g \perp \Dp,\: t \geq 0.$ Also we obtain easily $V(t)g \perp \Dm,\: t \geq 0.$ Indeed, for $h \in \Dm$ we have
$$(V(t)g, h) = (g, V^*(t)h) = (g, U_0(-t)h) = 0$$
since $U_0(-t)h \in \Dm$ for $t \geq 0.$
R. Phillips called the solutions with data $h \in {\mathcal D}(G_b)$ with $V(t)h \perp (\Dp \oplus \Dm),\: t \geq 0$ {\it incontrollable} (see \cite{Ma1}).\\

We modify the argument of Lemma 2.2 in \cite{G3} (see also Proposition 4.2.5 in \cite{P}) established for strictly hyperbolic systems in order to cover the situation when $A(\xi)$ has eigenvalues of finite multiplicity and $\Ker A(\xi)$ is not trivial.
Let $\varphi(x) \in C^{\infty}(\R^n)$ be such that $\varphi(x) = 1$ for $|x| \geq 2\rho,\: \varphi(x) = 0$ for $|x| \leq \rho.$ Set $w(t,x) = \varphi(x) V(t)f.$ We have
$$w(t, x) = G \varphi V(t) f_0 + [G, \varphi] V(t) f_0 = v(t, x) + [G, \varphi] V(t) f_0.$$
Since $[G, \varphi]$ has compact support, according to Proposition 3.1.9 in \cite{P}, we conclude that there exists a sequence $t_j \to +\infty$ such that $[G, \varphi] V(t_j)f_0 \longrightarrow 0$ as $t_j \to \infty.$ Thus by our hypothesis, we get $v(t_j, x) \to 0.$\\

It is cleat that  $v(t, .) \in H_0^{ac}$ and we may consider the translation representation of $v(t, .)$. On the other hand, $v(t,.) \in \bigcap_{j = 1}^{\infty}D(G_0^j)$ since $G_0 v = G_b v.$ Next we have
$$(\partial_t - G) v(t,x) =  \sum_{j = 1}^n (A_j \varphi_{x_j}) V(t)f = g(t,x).$$ 
Applying the transformation $\Rc_n$ to both sides of the above equality and setting 
$$h_j(s, \omega, t) = \langle \Rc_n(v)(s, \omega, t), r_j(\omega) \rangle,\: l_j(s, \omega, t) = \langle \Rc_n(g)(s, \omega, t), r_j(\omega) \rangle, \: j = 1,...,d,$$
we obtain the equations 
$$(\partial_t + \tau_j(\omega) \partial_s) h_j(s, \omega, t) = l_j(s, \omega, t),\: j =1,...,d.$$
Next we repeat the argument of Lemma 2.2 in \cite{G3} based on the following
\begin{lem} [\cite{G3}] Let $g \in \bigcap_{j=1}^{\infty} (G_0^j) \cap \Ha$ and let $\Rc_n g = 0$ for $|s| \geq b.$ Then $g = 0$ for $|x| \geq b$ if and only if
$$\int\int_{\R \times \sn} [A(\omega)]^k \big [ (\Rc_n g)(s, \omega) + (-1)^{(n-1)/2} (\Rc_n g)(-s, -\omega)\big ] s^a Y_j(\omega) ds d\omega = 0$$
for $a = 0,1,2,...$ and any spherical harmonic function $Y_m(\omega)$ of order $m \geq a + k +(3-n)/2.$

\end{lem}

Thus we conclude that $v(t,x) = 0$ for $|x| \geq 2\rho$, hence  $w(t, x) = 0$ for $|x| \geq 2 \rho$. We get $V(t)f = 0$ for $|x| > 2\rho$ and this completes the proof of Theorem 2.4.

\begin{rem} For systems with $\Ker A(\xi) = \{0\}$  V. Georgiev proved in \cite{G4} that if $f \in {\mathcal H}_{\infty}^{+} \cap (\Dm)^{\perp}$, then $V(t)f$ is a disappearing solution. On the other hand, the assumption $f \in (\Dm)^{\perp}$ cannot be relaxed. In fact, in section 3 we construct an  example of (ADS) for which $f \in \bigcap_{j=1}^{\infty} D(G_b^j) \cap {\mathcal H}_{\infty}^{+}$, but $f(x)$ has not compact support with respect to $x$.
\end{rem}
The space $\Dp$ is invariant with respect to the semigroup $V(t)$. Thus the generator $G_b$ is the extension of the generator $G_{+}$ of the group $V(t) \vert_{\Dp} = U_0(t)\vert_{\Dp}$. By using the translation representation $\Rc_n$, it is easy to see that $G_{+}$ has spectrum in $\{z \in \C: \Re z \leq 0\}$, so $\ii \R$ is in $\sigma(G_b)$
(see \cite{LP1}). On the other hand, it was proved in \cite{GS} that the eigenvalues of $G_b$ on $\ii \R$ have finite multiplicity and the only accumulating point of these eigenvalues  could be 0. Thus the $\sigma(G_b)$ is continuous on $\ii \R$. The analysis of $\sigma(G_b)$ in $\Re z < 0$ is more complicated. In the next section we show for the  Maxwell system that there exist eigenvalues $\lambda < 0$ of $G_b.$\\

To define a scattering operator we prove the existence of the operator
$$Wf = \lim_{t \to \infty}  U_0(-t) J V(t) f,\: f \in H_b^{\perp}$$
assuming the hypothesis (H) fulfilled (see \cite{GS}, \cite{P}). We define the scattering operator $S = W \circ W_-$ by using the diagram on Figure 1.

\begin{figure}[h!]
\centerline{\includegraphics*{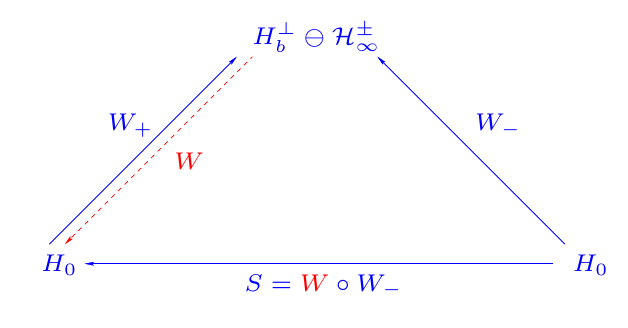}}
\caption{Scattering operator}
\end{figure}

\section{Asymptotically disappearing solutions for Maxwell system}
\renewcommand{\theequation}{\arabic{section}.\arabic{equation}}
\setcounter{equation}{0}

In this section we show that for the Maxwell system with maximal dissipative boundary conditions there exist (ADS). The Maxwell system in $\R^3$ is given by the equations
$$\partial_t E - \curl B = 0, \: \partial_t B + \curl E = 0.$$
$$\dive E = 0,\: \dive B = 0.$$

It is well known that the wave equation $u_{tt} - \Delta u = 0$ in $\R_t \times \R^3$ admit almost spherical solutions $\frac{f(|x| + t)}{|x|}$ defined outside $x = 0$. It was proved in \cite{CPR} that for Maxwell system there are no such almost spherical solutions with the  exception of functions $g(x) + ct$ linear in $t$.
To find a family of incoming divergence free solutions of Maxwell systems depending on $|x|$ and $t$, we apply the following

\begin{thm} [\cite{CPR}]
Let $h \in C^\infty(\R)$ and let $h^{(k)} = \partial^k_s h(s)\in L^1([0,\infty[)$, for all $k \in \N$.
Then
\begin{equation} \label{eq:3.1}
E \ :=\ 
\bigg(\frac{h''(|x|+t)}{|x|}
 -\frac{h'(|x|+t)}{|x|^2}
\bigg)\,
\frac{x}{|x|}
\:\wedge\:(1,0,0)\,,
\end{equation}
\begin{equation} \label{eq:3.2}
B\ :=\
-\bigg(
\frac{h''}{|x|}
-
\frac{3\,h^{\prime}}{|x|^2}
+
\frac{3\,h}{|x|^3}
\bigg)\ 
\frac{x}{|x|}
\wedge\bigg(
\frac{x}{|x|}
\wedge
\big(1,0,0\big)
\bigg)
\ +\
2\,\bigg(\frac{h'}{|x|^2}
- \frac{h}{|x|^3}
\bigg)\,
\big(1,0,0\big),
\end{equation}
where
the argument of the functions $h^{(k)}$ 
is $|x|+t$, define smooth divergence free incoming solutions
of Maxwell system in $\R_t^+ \times (\R^3 \setminus  0)$.
\end{thm}

The proof of this theorem is technical. The starting point is to search $E$ in the form $E = \curl (g, 0, 0)$ with 
$g(t, x) = \frac{f(|x| + t)}{|x|}$. This guarantees $\dive E = 0$ and $\Box E = 0.$ Next we determine $B$ from the equation $B_t = - \curl E$ so that $\dive B = 0.$ Finally,
$$ \partial_t(E_t - \curl B) = E_{tt} + \curl \curl E = E_{tt} - \Delta E = 0$$
and since $E_t - \curl B \to 0$ as $t \to \infty$, we obtain $E_t - \curl B = 0.$
 We can exchange the role of $E$ and $B$ and start with $B = \curl (g, 0, 0)$, but this leads to similar expressions.

We wish to construct asymptotically disappearing solutions 
in $|x|>1$ that satisfy
a homogeneous boundary condition
\begin{equation} \label{eq:3.3}
u=(E,B)\  \in \ {\mathcal N}(x),\quad
{\rm on}\quad |x|=1\,.
\end{equation}
Here $\cn(x)$ is a four dimensional linear subspace
of $\C^6$ depending smoothly of $x$.

Write the Maxwell equations in matrix form
$$ u_t 
\ -\ 
\sum_{j=1}^3
A_j\, \partial_j u
\ =\ 0\,.
$$
The matrices $A_j$ are real symmetric and 
$A(\xi) := \sum_{j = 1}^3 A_j\xi_j$ for $\xi \not= 0$ has rank equal to 4 and eigenvalues $0, \pm |\xi|$ of multiplicity 2. 

 A sufficient condition for $\cn(x)$ to be maximally dissipative and to obtain a well
posed mixed initial boundary value problem generating a 
contraction semigroup $V(t)$ on $(L^2(\{|x|\ge 1\}))^6$ is that
$$
{\rm dim}\, \cn(x) =4,
\quad
{\rm and},
\quad
\big\langle
A(\nu(x)) u\,,\, u\big\rangle \ \le \ 0, \: \forall u \in \cn(x), \: \forall |x| = 1.
$$
 It follows that $\cn(x)\supset {\rm Ker}\,A(\nu(x))$ for all boundary points
$x \in \partial \Omega$.

For any unit vector $\nu$ the eigenvalues of $A(\nu)$ are
$-1,0,1$.  The kernel of $A(\nu)$ is the set of $(E,B)$ so that  both
$E$ and $B$ are parallel to $\nu$.   The condition that
$\cn(x)$ contain the kernel is equivalent to say that $(E, B)$
belonging to $\cn(x)$ if $(E, B)$ is determined entirely by the tangential
components $(E_{tan},B_{tan})$.  The eigenspace $\Sigma_\pm(\nu)$ of $A(\nu)$ corresponding
to eigenvalue $\pm 1$ is equal to 
$$\Sigma_\pm(\nu) 
\ :=\ 
\big\{(E,B)\,:\,
E_{tan}\ =\ \mp\, \nu \wedge B_{tan}\big\}\,.
$$
The span of eigenspaces $\Sigma_{-}(\nu) \oplus \Ker A(\nu)$ with
non-positive eigenvalues  is strictly dissipative, that is  for all $u\in \Sigma_-(\nu) \oplus \Ker A(\nu)$ we have
 $$
 \big\langle
 A(\nu) u\,,\, u\big\rangle 
\ =\ 
-
\|u_{tan}\|^2
\ =\ 
 - \big\|( E_{tan}, B_{tan}) \big\|^2\,.$$
To construct asymptotically disappearing solutions we choose $h$
in a special way in Theorem 3.1.
\begin{thm} [\cite{CPR}] 
\label{thm:asympt} 
Let  $\epsilon_0 > 0$ be sufficiently small and
for  $0<\epsilon <\epsilon_0$ set 
$2r= 1- \sqrt{1 + 4/\epsilon}<0$ and $h(s)=e^{rs}$.  Then $(E,B)(t,x)$ defined by 
\eqref{eq:3.1} and \eqref{eq:3.2} yield a
divergence free solution
of boundary value problem defined
by the Maxwell equations in $|x| > 1$ with 
maximal dissipative boundary condition
\begin{equation} \label{eq:3.4}
(1 + \epsilon) E_{tan}  - \nu \wedge B_{tan}  = 0, \qquad {\rm on} \qquad |x|=1.
\end{equation}
For each $\alpha$  there is a constant $C(\epsilon, \alpha) > 0$ so that
$ \big|\partial^\alpha (E,B)(t,x)\big| \leq C(\epsilon, \alpha)\, h(t+|x|)$.  In particular, the energy decays exponentially as $t \to \infty$ and $G_b$ has an eigenvalues $r < 0$.
\end{thm}

It is important to note that for the mixed problem with boundary conditions (\ref{eq:3.4}) and $\epsilon > 0$ there are no disappearing solutions. This follows from Theorem 3 in \cite{G4} saying that if $\cn(x) \cap \Sigma_{-}(\nu(x)) = \{0\},\: \forall x \in \partial \Omega,$ for system with real analytic boundary conditions, then there are no disappearing solutions. In our situation, if $(E, B) \in \cn(x) \cap \Sigma_{-}(\nu(x)),$ we have $\epsilon E_{tan} = 0$ and this yields $B_{tan} = 0$, so $(E, B) \in \Ker A(\nu(x))$. On the other hand, it is clear that for the sphere $|x| = 1$ the boundary condition  (\ref{eq:3.4}) is real analytic with respect to $x$. For $\epsilon = 0$ the boundary condition
\begin{equation} \label{eq:3.5}
E_{tan}  - \nu \wedge B_{tan}  = 0, \qquad {\rm on} \qquad |x|=1
\end{equation}
satisfies $E_{-}(\nu(x)) \subset \cn(x),\: \forall x \in \partial \Omega$ and (\ref{eq:3.5}) is the analog of the condition $\Bigl((\partial_{\nu} + \partial_t)u\Bigr)\vert_{\partial \Omega} = 0$, for the wave equation $\partial_{tt} - \Delta u = 0$ (see \cite{Ma2} for the results concerning this mixed problem).\\

It is interesting to see that for the Maxwell system with dissipative boundary condition, if $G_b g = \lambda g$ with $\Re \lambda < 0$, then $g(x) = {\mathcal O} (e^{\Re \lambda |x|})$ as $|x| \to \infty.$ To see this, consider the function $\varphi(x)$ introduced in Section 2
and set $w(x) = \varphi(x) g(x).$ We have
$$G \Bigl( \begin{matrix} E \\ B \end{matrix}\Bigr) = \Bigl(\begin{matrix} 0 &  \curl \\
-\curl &  0\end{matrix} \Bigr) \Bigl( \begin{matrix} E \\ B \end{matrix}\Bigr).$$
Taking
$$Q\Bigl(\begin{matrix} E \\ B \end{matrix}\Bigr) = \Bigl(\begin{matrix} \dive E\\ \dive B \end{matrix} \Bigr),$$
we have $Q(\xi) A(\xi) = 0$, $\Ker Q(\xi) = \Im A(\xi)$
and we obtain $ (G^2 + Q^* Q)\Bigl(\begin{matrix} E \\ B \end{matrix}\Bigr) = \Delta \Bigl(\begin{matrix} E \\ B \end{matrix}\Bigr).$ Let $g = \Bigl(\begin{matrix} E \\ B \end{matrix}\Bigr)$ be an eigenfunction of $G_b$ with eigenvalue $\lambda, \Re \lambda < 0.$ Obviously, $G_b g = \lambda g$ implies  $Q g = 0$. Thus we get
$$(\Delta + (\ii\lambda)^2) w = [(G^2 + Q^* Q) - \lambda^2] w = (G^2_b - \lambda^2) w + Q^* \Bigl( \begin{matrix} \la \grad \varphi, E \ra \\ \la \grad \varphi, B \ra \end{matrix} \Bigr) $$
$$=  (G_b + \lambda) [G_b, \varphi] g+ Q^* \Bigl( \begin{matrix} \la \grad \varphi, E \ra \\ \la \grad \varphi, B \ra \end{matrix} \Bigr) = F_{\varphi} (g).$$
Here we have used the fact that $G^2 w = G_b^2 w$. The right-hand side $F_{\varphi} (g)$ has compact support and $\|F_{\varphi}(g)\| \leq C \|g\|$ with constant depending on $\varphi.$ The (incoming) resolvent of the free Laplacian $R_{-}(\mu) = (\Delta + \mu^2)^{-1}$ for $\Im \mu < 0$ has kernel
$$R_{-}(x, y; \mu) = -\frac{e^{-\ii \mu |x- y|}}{4 \pi |x-y|}$$ 
By using the above equation for  $w$ and the kernel of $R_{-}(\ii \lambda)$, we obtain
$$(\varphi g) (x)=  -\frac{1}{4 \pi} \int \frac{e^{\lambda |x - y|}}{|x - y|} F_{\varphi} (g)(y) dy$$
and this yields
$$|g(x)|\leq C_0 e^{(\Re \lambda) |x|} \|g\|, \:{\rm for}\: |x| \geq 3 \rho.$$

For the Maxwell system with strictly dissipative boundary conditions it is natural to conjecture that the spectrum of $G_b$ in $\{z \in \C:\: \Re z < 0\}$ is formed by isolated eigenvalues with finite multiplicity. This conjecture has been proved recently \cite{CPR1} for boundary problems satisfying the coercive conditions $(H)$ and $(H^*).$

\section{Representation of the scattering kernel}
\renewcommand{\theequation}{\arabic{section}.\arabic{equation}}
\setcounter{equation}{0}

In this section we obtain a representation of the scattering kernel. Such a representation has been obtained in \cite{MT} and  \cite{P2} for strictly hyperbolic systems with respectively conservative and dissipative boundary conditions. Here for completeness we sketch the proof of a representation in the case when $G$ has characteristics of constant multiplicity.  By using the translation representation $\Rc_n$ of $U_0(t)$, consider the scattering operator
$$\tilde{S}k = \Rc_n S \Rc_n^{-1}k = \lim_{t' \to +\infty} \Rc_n  U_0(-t')J V(2t')J^*U_0(-t')\Rc_n^{-1} k.$$
Let $k(s, \omega) \in (C_0^{\infty}(\R \times \sn))^d$ with $k = 0$ for $|s| > R_1$. Next for simplicity we write $\Rc$ instead of $\Rc_n.$  Set $f = \Rc^{-1} k$. We must study
$$ \lim_{t' \to +\infty} T_{-t'} \Rc \Bigl(J V(2t') J^* U_0(-t')f\Bigr).$$
Choose $p > R_1 + \frac{\rho}{v_0}.$ It is easy to see that for $t' > p$ we have $J^*U_0(-t')f = U_0(-t')f \in D_{-}^{\rho}$. Set $u_0(t, x) = U_0(t)f$ and denote by $u(t, x; t') = V(t)U_0(-t')f$ the solution of the problem
\begin{equation}
\begin{cases} (\partial_t - G)u = 0 \: {\rm in} \: \R^+ \times \Omega,\\
u(t, x)  \in {\mathcal N} (x) \: {\rm on}\: \R^+ \times \partial \Omega,\\
u\vert_{t \leq t' - p} = u_0(t- t', x).\end{cases}
\end{equation}

Consequently,  
\begin{equation} \label{eq:4.2}
\tilde{S}k = \lim_{t' \to +\infty} T_{-t'} \Rc J\tilde{u}(t', x),
\end{equation}
where $\tilde{u}(t,x)$ is the solution of the problem
\begin{equation} \label{eq:4.3}
\begin{cases} (\partial_t - G)\tilde{u} = 0 \: {\rm in} \: \R \times \Omega,\\
\tilde{u}(t, x)  \in {\mathcal N}(x) \: {\rm on}\: \R \times \partial \Omega,\\
\tilde{u}\vert_{t \leq - p} = u_0(t, x).\end{cases}
\end{equation}
Next we repeat the argument of Section 4 in \cite{P2}. Setting
$$v = \begin{cases} \tilde{u}\:\: {\rm in}\:\: \R \times (\R^n \setminus K),\\
0\:\:{\rm in}\:\: \R \times K, \end{cases}$$
we have
$$\la R((\partial_t - G)v), r_j(\omega) \ra = (\partial_t + \tau_j(\omega) \partial_s)\la Rv, r_j(\omega) \ra$$
and we get
\begin{equation} \label{eq:4.4}
\la Rv, r_j(\omega)\ra = \la Ru_0, r_j(\omega) \ra + \int_{-\infty}^t \la R((\partial_t- G)v)(\tau, s +\tau_j(\omega)(\tau- t), \omega), r_j(\omega)\ra d\tau.
\end{equation}
On the other hand, for every vector-valued function $\psi \in (C_0^{\infty}(\R^{n +1}))^d$ we obtain
$$((\partial_t - G)v, \psi) = \int \int_{\R \times \partial \Omega} \la v, A(\nu(x)) \psi \ra d\tau dS_x$$
and since $(\partial_t - G)v$ has compact support with respect to $x$, we can apply the above formula combined with (\ref{eq:4.4}) for the calculus of $(\Rc v)(s, \omega)$. Taking the limit $t' \to +\infty$ in (\ref{eq:4.2}), we deduce

$$(\tilde{S}k)(s, \theta) = k(s, \theta) + d_n \sum_{j=1}^d \tau_j(\theta)^{1/2} r_j(\theta) $$
$$ \times \int \int_{\R \times \partial \Omega} \delta^{(n-1)/2}(\la x, \theta \ra - \tau_j(\theta)(s + \tau))\big \la r_j(\theta), A(\nu(x)) \tilde{u}(\tau, x)\big \ra d\tau dS_x,$$
where $d_n$ is a constant depending only on $n$ and the integral is taken in the sense of distributions.

Next we repeat the argument of Section 3, \cite{MT} and \cite{P2}. Set $\tilde{u}(t,x) = u_0(t,x) + u^s(t,x),$ 
$$w_k^o (t, x, \omega) = \tau_k(\omega)^{1/2} \delta^{(n-1)/2}( \langle x, \omega \rangle - \tau_k(\omega)t) r_k(\omega),$$
and consider the (outgoing) solution $w_k^s(t, x, \omega)$ of the problem
\begin{equation}
\begin{cases} (\partial_t - G)w_k^s  = 0 \: {\rm in} \: \R \times \Omega,\\
w_k^s + w_k^o   \in {\mathcal N}(x) \: {\rm on}\: \R \times \partial \Omega,\\
w_k^s \vert_{t \leq - \frac{\rho}{v_0}} = 0.\end{cases}
\end{equation}
called disturbed plane wave.

To justify the existence of $w_k^s$, we set $w_k^s = z_k - w_k^o$ and consider the mixed problem
\begin{equation} \label{eq:3.1}
\begin{cases} (\partial_t - G)\tilde{z}_k  = 0 \: {\rm in} \: \R \times \Omega,\\
\tilde{z}_k  \in {\mathcal N}(x) \: {\rm on}\: \R \times \partial \Omega,\\
\tilde{z}_k\vert_{t \leq - \frac{\rho}{v_0}} = (-\tau_k(\omega))^{-(n+2)/2} H( \la x, \omega \ra - \tau_k(\omega) t) r_k(\omega),\end{cases}
\end{equation}
where 
$$H(\eta) = \begin{cases}  0,\:\: \eta >0,\\ \eta,\:\: \eta \leq 0\end{cases}.$$ 
It is easy to show the existence of $\tilde{z}_k$ and we get $z_k = \ii\partial_t^{(n+3)/2} \tilde{z}_k.$ By using $w^s_k + w^o_k,\: k = 1,...,d$, we can express $\tilde{u}(\tau, x)$ by $k(s, \omega)$ and we obtain a representation
$$(\tilde{S} k )(s, \theta) = k(s, \theta)  + \int\int_{\R \times \sn} K^{\#}(s - \tau, \theta, \omega) k(\tau, \omega) d\tau d\omega.$$
The distribution $K^{\#}(s, \theta, \omega) = \Bigl(S^{j k}(s, \theta, \omega)\Bigr)_{j,k = 1}^d$ is called scattering kernel. Thus we have the following
\begin{thm}   The scattering kernel $K^{\#}(s, \theta, \omega)$ computed with respect to the basis $\{r_j(\omega)\}_{j=1}^d$ has elements

$$S^{j k}(s, \theta, \omega) =  d_n^2 \tau_j(\theta)^{1/2}$$
\begin{equation} \label{eq:3.2}
\times \int\int_{\R \times \partial \Omega} \delta^{(n-1)/2}(\langle x, \theta \rangle - \tau_j(\theta)(s + t))\big\langle r_j(\theta), A(\nu(x))  (w_k^o + w_k^s)(t, x, \omega)\big\rangle dt dS_x.
\end{equation}
\end{thm}

\section{Back scattering inverse problem for the scattering kernel}
\renewcommand{\theequation}{\arabic{section}.\arabic{equation}}
\setcounter{equation}{0}
\def\Sb{{\mathbb S}}

 Consider the scattering operator $S$. It is easy to see that $(S - Id) (\Dm) \subset (\Dp)_0^{\perp}$, where $(. )_0^{\perp}$ denotes the orthogonal complement in $\Ha$. Then taking $k(\sigma, \omega) = 0$ for $\sigma > -\rho/v_0,$ we deduce $((\tilde{S} - I) k)(s, \theta) = 0$ for $s > \rho/v_0$ and we get 
$$K^{\#}(s, \theta, \omega) = 0\: {\rm for}\: s > 2\rho/v_0.$$
This property implies that the Fourier transform $\hat{K}^{\#}(\lambda, \theta, \omega)$ of $K^{\#}(s, \theta, \omega)$ admits an analytic continuation for $\Im \lambda < 0$ and the same is true for the operator-valued function $\tilde{S}(s): (L^2(\sn))^d \longrightarrow (L^2(\sn))^d$.\\

In fact, a more precise result holds for the back-scattering matrix $S^{j k}(s, -\omega, \omega)$.
\begin{thm} We have
\begin{equation} \label{eq:5.1}
\max_{s \in \R}\:{\rm supp}\: S^{j k}(s, -\omega, \omega) \leq -\Bigl(\frac{\tau_j(-\omega) + \tau_k(\omega)}{\tau_j(-\omega) \tau_k(\omega)}\Bigr) \rho(\omega),
\end{equation}
where $\rho(\omega) = \min_{y \in \partial \Omega} \la y, \omega \ra$ is the support function of $\partial \Omega$ in direction $\omega.$
\end{thm}

Applying (\ref{eq:4.7}), the proof of the above inequality is the same as that in Theorem 3.2 in \cite{MT}.\\

Before going to the analysis of an equality in (\ref{eq:5.1}), consider the problem  for the wave equation with dissipative boundary conditions
\begin{equation} \label{eq:5.2}
\begin{cases} (\partial_t^2 - \Delta) w = 0\: {\rm in} \: \R^+ \times \Omega,\\
\partial_{\nu} w + \gamma(x) \partial_t w = 0\:\: {\rm on}\: \R^+ \times \partial \Omega,\\
(w(0, x), w_t(0, x)) = (f_1, f_2),\end{cases}
\end{equation}
where $\gamma(x) \geq 0, x \in \partial \Omega$ is a smooth function.
We can introduce a scattering operator $S \in {\mathcal L}(L^2(\R \times \sn))$ related to (\ref{eq:5.2}) and the kernel $K^{\#}(s- s', \theta, \omega)$ of $S - Id$ is called
scattering kernel. The singularities of $K^{\#}(s, \theta, \omega)$ with respect to $s$ are closely related to the geometry of the obstacle. Thus if $\gamma(x) \neq 1, \: \forall x \in \partial \Omega,$ and $ \theta \neq \omega$ we have (see \cite{Ma3}) for strictly convex obstacles
\begin{equation} \nonumber
\max \:{\rm sing supp}_s\: K^{\#}(s, \theta, \omega)\} = \max_{y \in \partial \Omega} \langle y, \theta - \omega\rangle.
\end{equation}
For back scattering $\theta = -\omega$ this yields
\begin{equation} \label{eq:5.3}
\max \:{\rm sing supp}_s\: K^{\#}(s, -\omega, \omega) = -2 \rho(\omega) = -T_{\gamma}.
\end{equation}
Here $T_{\gamma}$ is the shortest sojourn time of a ray incoming with direction $\omega$ and outgoing with direction $-\omega$ (see \cite{PS} for the definition of sojourn time).
To obtain (\ref{eq:5.3}), it is necessary to study the asymptotics of the {\it filtered} scattering amplitude
$$a_{\varphi}(\lambda, -\omega, \omega) = \int e^{-\ii \lambda s} K^{\#}(s, -\omega, \omega) \varphi(s) ds,$$
where $\varphi(s) \in C_0^{\infty}(\R)$ has small support around $-T_{\gamma},\: \varphi(-T_{\gamma}) = 1$ and the integral is taken in the sense of distributions. By using the propagation of the wave front sets of the solutions of the mixed problem (\ref{eq:5.2}) in the diffraction region  and a microlocal parametrix, Majda \cite{Ma3} showed that if the support of $\varphi$ is sufficiently small, then
\begin{equation} \label{eq:5.4}
a_{\varphi}(\lambda, -\omega, \omega) = c_n \lambda^{(n-1)/2}{\mathcal K}^{-1/2}(x_+) e^{\ii\lambda T_{\gamma}} \Bigl(\frac{1 - \gamma(x_+)}{1 + \gamma(x_+)} + {\mathcal O}(|\lambda|^{-1})\Bigr).
\end{equation}
Here $x_+ \in \partial \Omega$ is the unique point on $\partial \Omega$ with $\nu(x_+) = \omega$ and ${\mathcal K}(x_+)$ is the Gauss curvature at $x_+.$ For $\gamma(x) \not= 1$  similar result holds for arbitrary (non-convex) obstacles \cite{Ma4} (see also \cite{P}) and there exists an open dense subset $\Sigma \subset \sn$  such that for every $\omega \in \Sigma$ we have
\begin{equation} \label{eq:5.5}
\max\: {\rm sing supp}_s\: K^{\#}(s, -\omega, \omega) = -2 \rho(\omega).
\end{equation}
Thus from the leading singularities of the back scattering kernel we can determine the convex hull of the obstacle.
If $\gamma(x) = 1, \: \forall x \in \partial \Omega$, the leading term in (\ref{eq:5.4}) vanishes for all directions $\omega \in \sn.$
In this direction
V. Georgiev and J. Arnaoudov, proved the following
\begin{thm} [\cite{GA}] Let $\gamma(x) = 1$ for all $x \in \partial \Omega.$ Then for $n = 3$ if $K$ is strictly convex, we have
$$a_{\varphi}(\lambda, -\omega, \omega) = \frac{c}{16\pi}{\mathcal K}^{-1/2}(x_+) e^{\ii\lambda T_{\gamma}} + {\mathcal O}(|\lambda|^{-1})$$
with $c \neq 0.$ Moreover, for an arbitrary smooth obstacle there exists an open dense subset $\Sigma \subset \sn$  such that for every $\omega \in \Sigma$ we have $(\ref{eq:5.5}).$
\end{thm}
\begin{rem} Let us note that it is not proved that the problem (\ref{eq:5.2}) with $\gamma(x) = 1,\: \forall x \in \partial \Omega,$ has disappearing solutions and this explains the inverse scattering result in the above theorem. In \cite{Ma1} Majda established the existence of disappearing solutions for mixed problem outside the sphere $|x| = 1$  with boundary condition  $\partial_{\nu} w + \partial_t w + w = 0$ on $\R \times {\mathbb S}^2.$

\end{rem}
Passing to the case of symmetric systems, consider first strictly hyperbolic systems with $\Ker A(\xi) = \{0\}$ and maximal energy preserving boundary condition 
$$\langle A(\nu(x)) u, u \rangle = 0,\: u \in \cn(x).$$
In this case $r/2 = d$ and  there exists (see Chapter VI, \cite{LP}) an orthonormal basis $(p_j(x))_{j=1}^{d}$ of the positive eigenspace $\Sigma_{+}(\nu(x))$ of $A(\nu(x))$ with respect to the inner product $< A(\nu(x))u, u >$ and an orthonormal basis $(n_j((x))_{j=1}^{d}$ of the negative eigenspace
$\Sigma_{-}(\nu(x))$ of $A(\nu(x))$ with respect to the inner product $-< A(\nu(x))u, u >.$ Then a maximal energy preserving space $\cn(x)$ is spanned by the vectors $(p_j(x) + n_j(x))_{j=1}^{d}$ and we have $u(x) \in \cn(x)$ if and only if
$$\langle p_j(x) + n_j(x), A(\nu(x))u(x) \rangle = 0,\: 1\leq j \leq d.$$

When we change $x$ on $\partial \Omega$, this special basis changes. For strictly hyperbolic systems with maximal energy preserving boundary conditions Majda and Taylor \cite{MT} assuming that $(p_j)$ and $(n_j)$ depend implicitly through the unit normal $\nu(x)$, proved that for every given $k, 1 \leq k \leq d,$ there exists an open dense
set $\Sigma \subset  \sn$ such that for every $\omega \in \Sigma$ for some $j,\: 1 \leq j \leq d,$ we have
\begin{equation} \label{eq:15}
\max  \: {\rm sing supp}_s\:  S^{jk}(s, -\omega, \omega) = - \Bigl(\frac{ \tau_j(-\omega) + \tau_k(\omega)}{\tau_j(-\omega) \tau_k(\omega)}\Bigr)\rho(\omega).
\end{equation} 

We should mention that for strictly convex obstacles Majda and Taylor showed \cite{MT} that the leading term of the filtered $(j,k)$ term of the scattering matrix has the form
$$c_n \lambda^{(n-1)/2} \exp\Bigl( \ii \lambda \Bigl(\frac{ \tau_j(-\omega) + \tau_k(\omega)}{\tau_j(-\omega) \tau_k(\omega)}\Bigr)\rho(\omega)\Bigr){\mathcal K}(x_{+})^{-1/2}$$
$$\times \sum_{\mu = 1}^d \la r_k(\omega), p_{\mu}(\omega)\ra \la r_j(-\omega), n_{\mu}(\omega)\ra.$$
When we fix $k, \: 1 \leq k \leq d$, the vector
$\sum_{\mu = 1}^{d} \la r_k(\omega), p_{\mu}(\omega)\ra n_{\mu}(\omega)$ is in the eigenspace $\Sigma_{-}(\omega)$ and it is easy to
see that for some $j,\: 1 \leq j \leq d,$ we have
$$\sum_{\mu = 1}^d \la r_k(\omega), p_{\mu}(\omega)\ra \la r_j(-\omega), n_{\mu}(\omega)\ra \neq 0.$$

In the case of maximal dissipative boundary conditions, the structure of the boundary space $\cn(x)$ is more complicated. It was shown \cite{P2}, \cite{GA1}  that we have the following
\begin{lem} Let $\cn(x)$ be a maximal dissipative linear space depending smoothly on $x \in \partial \Omega.$ 
Then for every $\hat{x} \in \partial \Omega$ there exists a neighborhood ${\mathcal V}$ of $\hat{x}$ and smooth in ${\mathcal V} \cap \partial \Omega$  vectors $p_1(x),...,p_d(x),n_1(x),...,n_{d}(x)$ satisfying
$$\la p_i(x), A(\nu(x)) p_j(x) \ra = \delta_{i,j},\: \la n_i(x), A(\nu(x)) n_j(x) \ra = - \delta_{i,j},$$
$$\la n_i(x), A(\nu(x))p_j(x) \ra = 0,\: i,j = 1,...,d,$$
and $0 \leq \mu(x) \leq d$ so that $\cn(x) \ominus (\Ker A(\nu(x)))$ is spanned by the vectors
$$\{p_1(x) + n_1(x),...,p_{\mu(x)}(x) + n_{\mu(x)}(x), n_{\mu(x) + 1}(x),...,n_{d}(x)\}.$$
\end{lem}
In the strictly dissipative case we have $\mu(x) = 0.$  It is important to note that in general $(p_j(x))_{j=1}^d$ and $(n_j(x))_{j=1}^d$ do not form a basis, respectively in the spaces $\Sigma_{+}(\nu(x))$ and  $\Sigma_{-}(\nu(x)).$ On the other hand, $u(x) \in \cn(x)$ if and only if
\begin{equation} \label{eq:5.7}
\begin{cases}\la p_j(x) + n_j(x), A(\nu(x)) u(x)\ra = 0,\: j =1,...,\mu(x),\\
\la p_j(x), A(\nu(x))u(x) \ra = 0,\: j = \mu(x) +1,...,d.\end{cases}
\end{equation}
It is clear that for strictly dissipative boundary conditions if 
\begin{equation} \label{eq:5.8}
 \cn(x) \ominus (\Ker A(\nu(x)) = \Sigma_{-}(\nu(x)),
\end{equation}
then $(n_j(x))_{j= 1}^d$ span $\Sigma_{-}(\nu(x))$ and $(p_j(x))_{j=1}^d$ span $\Sigma_{+}(\nu(x)).$
The condition (\ref{eq:5.8}) is the analog of the boundary condition (\ref{eq:5.2}) with $\gamma(x) = 1, \: \forall x \in \partial \Omega.$

For strictly hyperbolic systems we have the following

\begin{thm} [\cite{P2}] Let $A(\xi)$ have simply characteristic roots for $\xi \not= 0$ and let $\Ker A(\xi) = \{0\}.$ Consider the problem $(\ref{eq:1.1})$  and assume that the vectors $(p_j)$ and $(n_j)$ depend implicitly through the unit normal $\nu(x)$ and $(p_j)$ and $(n_j)$ are smooth functions of this normal with $\mu$ independent on $\nu(x).$  Moreover, assume that for every $x \in \partial \Omega$ in Lemma $5.4$ $\mu(x) = const$ and if $\mu = 0$, we have
\begin{equation} \label{eq:5.9}
 \cn(x) \ominus (\Ker A(\nu(x)) \not= \Sigma_{-}(\nu(x)).
\end{equation}
 Then there exists an open dense
set $\Sigma \subset \sn$ such that for every $\omega \in \Sigma$ there exist $(k, j)$ depending on $\omega$ so that we have
\begin{equation} \label{eq:5.10}
\max  \: {\rm sing supp}_s\:  S^{jk}(s, -\omega, \omega) = - \Bigl(\frac{ \tau_j(-\omega) + \tau_k(\omega)}{\tau_j(-\omega) \tau_k(\omega)}\Bigr)\rho(\omega).
\end{equation} 
\end{thm}

\begin{rem} In contrast to the problem (\ref{eq:5.2}) with $\gamma(x) \not= 1,\: \forall x \in \partial \Omega$, the condition (\ref{eq:5.8}) does not exclude the existence of disappearing solutions. On the other hand, we are interesting to have at least one term in the scattering matrix $S^{jk}(s, -\omega, \omega)$ with leading singularity given by (\ref{eq:5.10}). This means that all other terms could be regular or vanishing.\\

In the proof of Theorem 5.5 the crucial point is the construction of a microlocal outgoing parametrix with boundary data the
distribution $\tau_k(\omega) \delta^{(n-1)/2}(\langle x, \omega \rangle - \tau_k(\omega)t)r_k(\omega)$. More precisely, we assume that locally the boundary is given by $x = 0$ with $(x, y) \in \R^+ \times \R^{n-1}$. After a microlocalization around the point $(0, y, t, 0, -1)$ with $\nu(y) = \omega$ and an application of the decoupling procedure of the matrix symbol \cite{T}, we are going to study the problem (see \cite{MT}, \cite{P2})
$$ \frac{\partial}{\partial x} \tilde{v}_k - \begin{bmatrix}\lambda_1 & & 0\\ & \ddots & \\ 0  & &\lambda_r\end{bmatrix}\tilde{v}_k = F(x, y, t),$$
\begin{equation} \label{eq:5.11}
\Lambda \Bigl((V \tilde{v}_k)\vert_{x = 0} - \Bigl(\tau_k(\omega)^{1/2} \delta^{(n-1)/2} (\la x, \omega \ra - \tau_k(\omega) t) r_k(\omega)\Bigr)\vert_{\partial \Omega}\Bigr) = g(y,t),
\end{equation}
$$\tilde{v}_k \:  {\rm is}\: {\rm smooth}\: {\rm for}\: t \leq - T_0 < 0.$$
Here $F(x, y,t),\: g(y, t)$ are smooth functions, $\lambda_j(x, y, t, \eta, \tau),\: j = 1,...,r,$ are different first order pseudodifferential operators, $V$ is a classical matrix pseudodiferential elliptic operator of order 0 and $\Lambda W(x) = 0$ means that $W \in \cn(x).$ For this construction we must determine the boundary data of $(V \tilde{v}_k)$ on $x = 0$ from (\ref{eq:5.11}). For this purpose, by using (\ref{eq:5.7}) and the condition (\ref{eq:5.9}) we obtain an elliptic  pseudodifferential system ${\mathcal E} \Bigl((V \tilde{v}_k)\vert_{x = 0}\Bigr) = Y_k$ on the boundary modulo smooth terms and choosing suitably $k$, we arrange $Y_k \not= 0.$ 
 Thus we determine the boundary data for $V \tilde{v}_k$ and we construct an outgoing paramerix in a standard way.
 
\begin{rem} Notice that in the case of the boundary condition (\ref{eq:5.8}), we get $Y_k = 0, k = 1,...,d,$ and the leading term of the back scattering matrix vanishes for all $j, k.$ 
\end{rem}
 We conjecture that the statement of Theorem 5.5 holds for symmetric systems with characteristics of constant multiplicity but this needs  extra work concerning the microlocal matrix reduction. For other inverse scattering problems for symmetric systems with dissipative boundary conditions we refer to \cite{G1}, where the case
of directions $(\theta, \omega) \in \sn \times \sn$ satisfying $\|\theta + \omega\| < \delta$ with sufficiently small $\delta > 0$ has been studied.

\end{rem}

\end{document}